\documentclass[prd,aps,showpacs,epsf,floats]{revtex4}%
\usepackage{amssymb}
\usepackage{amsfonts}
\usepackage{amsmath}
\usepackage{graphicx}%
\providecommand{\U}[1]{\protect\rule{.1in}{.1in}}
\begin{document}
\title{\textbf{An entropic analysis of approximate quantum error correction}}
\author{Carlo Cafaro$^{1,2}$ and Peter van Loock$^{2}$}
\affiliation{$^{1}$Max-Planck Institute for the Science of Light, 91058 Erlangen, Germany }
\affiliation{$^{2}$Institute of Physics, Johannes-Gutenberg University Mainz, 55128 Mainz, Germany}

\begin{abstract}
The concept of entropy and the correct application of the Second Law of
thermodynamics are essential in order to understand the reason why quantum
error correction is thermodynamically possible and no violation of the Second
Law occurs during its execution.

We report in this work our first steps towards an entropic analysis extended
to approximate quantum error correction (QEC). Special emphasis is devoted to
the link among quantum state discrimination (QSD), quantum information gain,
and quantum error correction in both the exact and approximate QEC scenarios.

\end{abstract}

\maketitle

\section{Introduction}

It is well-established that the notion of entropy plays a key role in the
foundations of quantum theory \cite{popescu, ferdy, renato, caticha, winter,
beretta} whose statistical nature is evident when dealing with incomplete
information gathered in quantum measurements. Incomplete information refers to
the fact that in quantum physics, as opposed to classical physics, two
non-commuting observables do not have any definite values simultaneously and
therefore one cannot obtain simultaneously perfect information about both. The
quantum mechanical perfect information gain always refers only to a complete
set of commuting observables. In fact, combining this aspect of quantum
mechanics with the notion of entanglement and nonlocality made Einstein,
Podolsky, and Rosen conclude that quantum mechanics is incomplete
\cite{einstein}. In general, measurements are performed to increase
information about physical systems. This information, if appropriate, may in
principle be used for a reduction of the thermodynamical entropy of such
physical systems.

In his 1929 seminal paper \cite{szilard}, Szilard presented the so-called
Szilard's engine to show that additional information about a system yields a
decrease in the entropy of the system. Szilard reaffirmed his belief in the
Second Law of thermodynamics and that the measurement process performed by
some sort of intelligent being (Maxwell's demon), in some overall sense,
requires energy dissipation. Szilard, however, did not pin down the exact
source of dissipation within a measurement cycle. In 1961, Landauer showed
that any erasure of information is accompanied by an appropriate increase in
entropy \cite{landauer1}. In 1982, relying on Landauer's key observations,
Bennett exorcised Maxwell's demon in a Szilard-like set-up \cite{bennett}.
Bennett's main conclusion was that the increase in entropy is not necessarily
a consequence of observations made by the demon, but accompanies the resetting
of the final state of the demon to be able to start a new cycle. In other
words, information gained has to eventually be erased, which leads to an
increase of entropy in the environment and prevents the Second Law of
thermodynamics from being violated. In fact, the entropy increase in erasure
has to be at least as large as the initial information gain. Bennett's
analysis was, however, completely classical. In 1984, Zurek analyzed the demon
quantum mechanically confirming Bennett's results \cite{zurek}. 

Perhaps, quantum error correction (QEC) is the best arena for considering the
links between entropy,\textbf{ }information\textbf{, }and thermodynamics. A
QEC technique consists in encoding quantum information into a physical system
in such a way that it can be either actively or passively saved from
decoherence \cite{charlie}. Furthermore, since the process of quantum
measurement cannot perfectly discriminate among non-orthogonal states, the
optimal strategy to encode information is to prepare the $d$-level quantum
system in one out of $d$ orthogonal states.

In this article, we discuss an additional application of Landauer's erasure
principle to show that quantum error correction, regarded as a Maxwell demon,
does not violate the Second Law of thermodynamics. The main initial motivation
for this work was the will of gaining a better understanding of the following
statement that appeared in \cite{vedral}: \textit{Doing perfect error
correction without perfect information gain is forbidden by the Second Law of
thermodynamics via Landauer's principle. This is analogous to von Neumann's
(1952) proof that being able to distinguish perfectly between two
non-orthogonal states would lead directly to the violation of the Second Law
of thermodynamics}.

The layout of this work is as follows. In Section II, we mention the main
historical objections to Landauer's principle which plays a key role in the
comprehension of the reason why QEC is compatible with the Second Law. In
Section III, we reconsider the standard entropic analysis of a QEC cycle
showing the compatibility of QEC with the Second Law. In Section IV, we
specify the meaning of exact- and approximate-QEC. Motivated by the aim of a
better understanding of Vedral's above-mentioned\textbf{ }statement, in
Section V we discuss the possibility of approximate-QEC where only an
imperfect discrimination of non-orthogonal quantum states is permitted\textbf{
}and underline some consequences of the presence of non-orthogonal quantum
states in the entropic analysis of a QEC cycle. Our final remarks appear in
Section VI.

\section{Brief historical background}

In his 1961 classic paper \cite{landauer1}, Landauer discussed the limitation
of the efficiency of computers imposed by physical laws. In particular, he
provided key arguments to solve Maxwell's demon puzzle in Szilard's engine.
Landauer's principle of information erasure states that when erasing one bit
of information stored in a memory device, on average, at least $k_{B}T\log2$
energy in the form of heat is dissipated into the environment. The quantity
$k_{B}$ denotes Boltzmann's constant while $T$ is the temperature of the
environment at which one erases. We stress that implicit in Landauer's
argument is the crucial assumption that information entropy translates into
thermodynamical entropy. Landauer's principle received several objections:

\begin{itemize}
\item The identification of information entropy with thermodynamical entropy
is unfounded \cite{jauch}. In particular, information gain should not be
identified with entropy decrease;

\item Landauer's claim is based only on the Second Law of thermodynamics and,
although plausible, not very rigorous. For instance, piston fluctuations
should be taken into consideration since they are of crucial importance in the
analysis of a Szilard engine \cite{berger};

\item Landauer's principle has no general validity since there exists a
superconducting logic device (the so-called quantum flux parametron) capable
of carrying out logically irreversible operations (information destruction,
for instance) without requiring any minimal dissipation per step \cite{goto}.
\end{itemize}

All these objections have been one by one rebutted to a certain extent. For
instance, the first objection was rebutted by Costa de Beauregard and Tribus
\cite{tribus}. They stress that the concept of entropy in statistical
mechanics can be deduced from the concept of information. The first objection
was also reconsidered later by Peres \cite{peres1, peres2} who, relying on
previous works of von Neumann \cite{john} (in 1952 von Neumann showed that
allowing for the possibility of distinguishing perfectly two non-orthogonal
quantum states would lead directly to the violation of the Second Law) and
Partovi \cite{partovi} (thermodynamic behavior is already present at the
quantum level and is not the exclusive domain of macroscopic systems),
concludes that there should be no doubt that entropy, as defined by von
Neumann in quantum theory and by Shannon \cite{shannon} in information theory
is fully equivalent to that of classical thermodynamics. However, we remark
that while entropy is measured in units of bits in classical information
theory, it is measured in units of joules/kelvin in classical thermodynamics.
This statement, however, he emphasizes, must be understood with the same vague
meaning as when we say that quantum notions of energy, momentum, angular
momentum, etc. are equivalent to the classical notions bearing the same names.
The second objection was addressed by Zurek \cite{zurek} who refined Szilard's
analysis by taking into consideration fully quantum aspects of Slizard's
engine. Finally, the third objection was considered by Landauer himself
\cite{landauer2} who stated that what was actually showed by Goto and
coworkers in \cite{goto} is that there is no minimal dissipation per step for
logically reversible operations and that this, in turn, does not contradict
his principle.

For a more detailed discussion about the objections to Landauer's principle,
we refer to Bennett \cite{bennett} who remarks that although Landauer's
principle in a sense is indeed a straightforward consequence or restatement of
the Second Law, it still has considerable pedagogic and explanatory power. It
makes clear that information processing and acquisition have no intrinsic
irreducible thermodynamic cost whereas the seemingly humble act of information
destruction does have a cost, exactly sufficient to save the Second Law from
the demon.

As a side remark, we emphasize that the Second Law is often regarded as being
statistical in nature: it can be violated in particular instances but not on
average. However, using tools from single-shot information theory, it was
shown in \cite{ferdy} that it can be applied to single systems as well.

Entropy and the Second Law are essential tools for a correct understanding of
the reason why QEC is thermodynamically possible and no violation of the
Second Law occurs during its execution \cite{nielsen1, nielsen2, vedral}. This
will be discussed in the next section.

\section{Entropic analysis of exact quantum error correction}

We follow the analysis presented in \cite{nielsen2}. From a thermodynamical
point of view, QEC may be regarded as a refrigeration process capable of
maintaining the quantum system at a constant entropy despite the environmental
noisy process whose tendency is to change the entropy of the quantum system
itself. Information about the quantum system gathered in quantum measurements
is used to keep the system cool. At first sight, it may actually appear that
QEC\ allows a reduction in the entropy of the quantum system in apparent
violation of the Second Law. However, a careful thermodynamic analysis shows
that QEC, like Maxwell's demon, does not violate the Second Law.

Consider a quantum system $\mathcal{Q}$ that is initially in the state $\rho$
with von Neumann entropy $\mathcal{S}\left(  \rho\right)  \overset{\text{def}%
}{=}-$Tr$\left(  \rho\log\rho\right)  $. The interaction of $\mathcal{Q}$ with
a noisy environment $\mathcal{E}$ takes generally $\mathcal{Q}$ to a new state
$\rho^{\prime}$ with entropy $\mathcal{S}\left(  \rho^{\prime}\right)
>\mathcal{S}\left(  \rho\right)  $. Ideally, when an exact-QEC (for the
meaning of exact- and approximate-QEC, we refer to Section IV\textbf{)} scheme
can be employed, the state $\rho^{\prime}$ with $\mathcal{S}\left(
\rho^{\prime}\right)  $ can return to $\rho$ with $\mathcal{S}\left(
\rho\right)  $. Thus, considering the entropy change of the system
$\mathcal{Q}$ just before (when the environmental noise has already acted upon
the quantum system of interest $\mathcal{Q}$) and right after QEC, one
concludes that%
\begin{equation}
\Delta\mathcal{S}\overset{\text{def}}{=}\mathcal{S}\left(  \rho\right)
-\mathcal{S}\left(  \rho^{\prime}\right)  <0\text{.} \label{ass}%
\end{equation}
From Eq. (\ref{ass}) it may appear that QEC violates the Second Law since
there is a reduction in entropy of $\mathcal{Q}$ (the total entropy of a
closed physical system cannot decrease). However, this is not the case as it
turns out from a proper thermodynamical analysis embracing \textit{all\ bodies
taking part in the process }($\mathcal{Q}$ is not a closed system)\textit{. }

First, assume that the quantum system of interest $\mathcal{Q}$ is in the
initial state $\rho$. After undergoing a noisy quantum evolution with a noisy
environment $\mathcal{E}$, the new state of $\mathcal{Q}$ becomes
$\rho^{\prime}$. We take into consideration the case in which $\mathcal{S}%
\left(  \rho^{\prime}\right)  >\mathcal{S}\left(  \rho\right)  $. As an
illustrative example concerning this last statement, consider a non-maximally
mixed quantum state $\rho\overset{\text{def}}{=}2^{-1}\left[  \left\vert
0\right\rangle \left\langle 0\right\vert +a\left\vert 0\right\rangle
\left\langle 1\right\vert +a\left\vert 1\right\rangle \left\langle
0\right\vert +\left\vert 1\right\rangle \left\langle 1\right\vert \right]
$\textbf{ }with $0\leq a\leq1$. Assuming an amplitude damping noise channel
$\Lambda_{AD}$ with damping parameter $\gamma\ll1$ \cite{nielsen2}, a simple
numerical calculation shows that $\mathcal{S}\left(  \Lambda_{AD}\left(
\rho\right)  \right)  \equiv\mathcal{S}\left(  \rho^{\prime}\right)
>\mathcal{S}\left(  \rho\right)  \simeq0.56$ provided that $0\leq
\gamma\lesssim0.25$. As a side remark, we point out that the amplitude damping
channel can be realized via a Jaynes-Cummings model, a theoretical model which
was originally used to study the classical aspects of spontaneous emission
\cite{jc}. We emphasize that the entropy\textbf{ }$\mathcal{S}\left(
\rho^{\prime}\right)  $\textbf{ }of the system in the final state\textbf{
}$\rho^{\prime}$\textbf{ }after the noisy quantum evolution can be less than
(or, equal to) the entropy\textbf{ }$\mathcal{S}\left(  \rho\right)  $\textbf{
}\cite{nielsen2}. For instance, it turns out that \cite{ben}: the class of
depolarizing channels causes entropy to increase for all states until it
reaches the maximum for the completely mixed state; for the dephasing class of
channels, entropy is nondecreasing: for some states it remains unchanged, and
for some states it increases; finally, for the amplitude damping class of
channels, entropy can decrease under the channel.

Second, a demon $\mathcal{D}$ (that is, an apparatus) carries on a syndrome
measurement on the state $\rho^{\prime}$ characterized by the measurement
operators $\left\{  \mathcal{M}_{k}\right\}  $. He obtains result $k$ with
probability $p_{k}$ and posterior state $\rho_{k}^{\prime}$ where,%
\begin{equation}
\rho_{k}^{\prime}\overset{\text{def}}{=}\frac{\mathcal{M}_{k}\rho^{\prime
}\mathcal{M}_{k}^{\dagger}}{p_{k}}\text{ and, }p_{k}\overset{\text{def}}%
{=}\text{Tr}\left(  \mathcal{M}_{k}\rho^{\prime}\mathcal{M}_{k}^{\dagger
}\right)  \text{.}%
\end{equation}
Third, the demon applies a unitary recovery operation $\mathcal{U}_{k}$ that
leads to the final state $\rho_{k}^{\prime\prime}$,%
\begin{equation}
\rho_{k}^{\prime\prime}\overset{\text{def}}{=}\mathcal{U}_{k}\rho_{k}^{\prime
}\mathcal{U}_{k}^{\dagger}=\frac{\mathcal{U}_{k}\mathcal{M}_{k}\rho^{\prime
}\mathcal{M}_{k}^{\dagger}\mathcal{U}_{k}^{\dagger}}{p_{k}}\text{.}%
\end{equation}
Finally, in order to regard this error correction procedure as a successful
cycle, it must be $\rho_{k}^{\prime\prime}=\rho$ for each measurement outcome
$k$. The cycle is then restarted. However, before restarting the cycle, the
demon $\mathcal{D}$ has to reset its (finite) memory. In other words, the
demon has to erase its record of the measurement result $k$. We shall see that
this fact causes an entropy production in the environment $\mathcal{E}$ (by
Landauer's principle) which is at least as large as the entropy reduction in
the quantum system $\mathcal{Q}$ being error corrected. We stress that the use
of the environment is essential for the erasure process. Without the coupling
of the memory device to the environment, it would be impossible to reset the
memory since no unitary evolution can transform a maximally mixed state with
entropy $\log2$ into a pure state with zero entropy \cite{peres2}.

This entropic analysis for the QEC cycle can be described in the following
terms. Recall that the initial state of the system $\mathcal{Q}$ is $\rho$.
After interacting with the noisy environment $\mathcal{E}$, its new state
becomes $\rho^{\prime}$. Thus, before performing the QEC procedure, system
$\mathcal{Q}$ is characterized by the state $\rho^{\prime}$. After exact-QEC ,
the state of $\mathcal{Q}$ is returned to $\rho$. Therefore, the net change in
entropy of the system $\mathcal{Q}$ due to error correction is $\Delta
\mathcal{S}$ in Eq. (\ref{ass}). However, as pointed out earlier, there is an
additional entropy cost associated with erasing the demon measurement record.
To reset its memory for the next QEC cycle, the demon must erase its
measurement record. This, in turn, causes a net increase in the entropy of the
environment $\mathcal{E}$ as prescribed by Landauer's principle. The number of
bits that must be erased is determined by the representation the demon
$\mathcal{D}$ uses to store the measurement result $k$. By Shannon's noiseless
channel coding theorem \cite{nielsen2}, at least $\mathcal{H}\left(
p_{k}\right)  $ bits are required on average to store the measurement result
($\mathcal{H}$ denotes Shannon's entropy function). Thus, a single QEC cycle
on average involves the dissipation of $\mathcal{H}\left(  p_{k}\right)  $
bits of entropy into the environment when the measurement record is
erased.\textbf{ }In summary, the total entropic cost for a single QEC cycle is
given by,%
\begin{equation}
\Delta\mathcal{S}_{tot}\overset{\text{def}}{=}\Delta\mathcal{S+}%
\mathcal{H}\left(  p_{k}\right)  =\mathcal{S}\left(  \rho\right)
-\mathcal{S}\left(  \rho^{\prime}\right)  +\mathcal{H}\left(  p_{k}\right)
\text{.}%
\end{equation}
To demonstrate that the Second Law is not violated, we need to show that
$\Delta\mathcal{S}_{tot}\geq0$. Let $\Lambda$ denote the noise process
occurring during the beginning of the QEC cycle such that $\Lambda\left(
\rho\right)  \overset{\text{def}}{=}\rho^{\prime}$. Furthermore, let
$\mathcal{R}$ be the quantum operation representing the error-correction
operation,%
\begin{equation}
\mathcal{R}\left(  \sigma\right)  \overset{\text{def}}{=}\sum_{k}%
\mathcal{U}_{k}\mathcal{M}_{k}\sigma\mathcal{M}_{k}^{\dagger}\mathcal{U}%
_{k}^{\dagger}\text{.}%
\end{equation}
As a side remark, we point out that perfect reversibility is obtained when%
\begin{equation}
\left(  \mathcal{R\circ}\Lambda\right)  \left(  \rho\right)  \equiv
\mathcal{R}\left(  \rho^{\prime}\right)  =\sum_{k}\mathcal{U}_{k}%
\mathcal{M}_{k}\rho^{\prime}\mathcal{M}_{k}^{\dagger}\mathcal{U}_{k}^{\dagger
}=\sum_{k}p_{k}\rho_{k}^{\prime\prime}=\left(  \sum_{k}p_{k}\right)  \rho
=\rho\text{,}%
\end{equation}
which holds true provided that $\rho_{k}^{\prime\prime}=\rho$, $\forall k$. We
also underline that such perfect reversibility is achieved provided we
consider the operator-sum decomposition of the map\textbf{ }$\Lambda$\textbf{
}restricted to the set of correctable errors only. The entropy introduced by
the error-correction operation $\mathcal{R}$ (acting on $\rho^{\prime}$) on
the environment $\mathcal{E}$ is quantified by the entropy exchange
$\mathcal{S}\left(  \rho^{\prime}\text{, }\mathcal{R}\right)  =\mathcal{S}%
\left(  \mathcal{W}\right)  $,%
\begin{equation}
\mathcal{S}\left(  \mathcal{W}\right)  \overset{\text{def}}{=}-Tr\left(
\mathcal{W}\log\mathcal{W}\right)  \text{,}%
\end{equation}
where the $\mathcal{W}$-matrix has elements $\mathcal{W}_{ij}$ defined as
$\mathcal{W}_{ij}\overset{\text{def}}{=}$Tr$\left(  \mathcal{U}_{i}%
\mathcal{M}_{i}\rho^{\prime}\mathcal{M}_{j}^{\dagger}\mathcal{U}_{j}^{\dagger
}\right)  $ \cite{nielsen2}. Observe that the diagonal elements of
$\mathcal{W}$ with $i=j\equiv k$ read,%
\begin{equation}
\mathcal{W}_{kk}\overset{\text{def}}{=}\text{Tr}\left(  \mathcal{U}%
_{k}\mathcal{M}_{k}\rho^{\prime}\mathcal{M}_{k}^{\dagger}\mathcal{U}%
_{k}^{\dagger}\right)  =\text{Tr}\left(  \mathcal{U}_{k}^{\dagger}%
\mathcal{U}_{k}\mathcal{M}_{k}\rho^{\prime}\mathcal{M}_{k}^{\dagger}\right)
=\text{Tr}\left(  \mathcal{M}_{k}\rho^{\prime}\mathcal{M}_{k}^{\dagger
}\right)  \overset{\text{def}}{=}p_{k}\text{.} \label{1}%
\end{equation}
Thus, the diagonal elements of $\mathcal{W}$ equal the probability $p_{k}$,
namely the probability that the demon $\mathcal{D}$ obtains measurement
outcome $k$ when measuring the error syndrome. Recalling that projective
measurements increase entropy, we obtain that the entropy of diagonal elements
of $\mathcal{W}$ is at least as great as the entropy of $\mathcal{W}$,
$\mathcal{S}\left(  \mathcal{W}_{kk}\right)  \geq\mathcal{S}\left(
\mathcal{W}\right)  $. However, because of Eq. (\ref{1}), we get that
$\mathcal{S}\left(  \mathcal{W}_{kk}\right)  =\mathcal{H}\left(  p_{k}\right)
$. Thus,%
\begin{equation}
\mathcal{H}\left(  p_{k}\right)  \geq\mathcal{S}\left(  \mathcal{W}\right)
=\mathcal{S}\left(  \rho^{\prime}\text{, }\mathcal{R}\right)  \text{.}
\label{2}%
\end{equation}
Observe that the equality in Eq. (\ref{2}) holds iff the off-diagonal terms in
$\mathcal{W}$ vanish, that is iff $\left\{  \mathcal{U}_{k}\mathcal{M}%
_{k}\right\}  $ form a canonical decomposition of $\mathcal{R}$ with respect
to $\rho^{\prime}$. Applying the subadditivity inequality for von Neumann
entropy \cite{nielsen2} to the joint system $\mathcal{Q}^{\prime}%
\mathcal{E}^{\prime}$ where $\mathcal{Q}^{\prime}$ is the quantum system of
interest after both the noise and the quantum correction have occurred
($\rho_{\mathcal{Q}^{\prime}}=\rho=\left(  \mathcal{R\circ}\Lambda\right)
\left(  \rho\right)  $) while $\mathcal{E}^{\prime}$ is the state of the
environment after the mentioned processes, it turns out that,%
\begin{equation}
\mathcal{S(Q}^{\prime}\text{, }\mathcal{E}^{\prime})\leq\mathcal{S(Q}^{\prime
})+\mathcal{S(E}^{\prime})=\mathcal{S(R}\left(  \mathcal{\rho}^{\prime
}\right)  )+\mathcal{S}\left(  \rho^{\prime}\text{, }\mathcal{R}\right)
\equiv\mathcal{S(\rho})+\mathcal{S}\left(  \rho^{\prime}\text{, }%
\mathcal{R}\right)  \text{.} \label{3}%
\end{equation}
Moreover, we have%
\begin{equation}
\mathcal{S(Q}^{\prime}\text{, }\mathcal{E}^{\prime})=\mathcal{S(}R^{\prime
})=\mathcal{S(}R)=\mathcal{S(Q})\equiv\mathcal{S}\left(  \rho^{\prime}\right)
\text{,} \label{4}%
\end{equation}
where $R^{\prime}$ and $R$ are the reference systems which purify
$\mathcal{Q}$ (the initial quantum system of interest before QEC) after and
before error correction, respectively. Thus, from Eqs. (\ref{3}) and
(\ref{4}), we obtain%
\begin{equation}
\mathcal{S}\left(  \rho\right)  -\mathcal{S}\left(  \rho^{\prime}\right)
+\mathcal{S}\left(  \rho^{\prime}\text{, }\mathcal{R}\right)  \equiv
\Delta\mathcal{S+S}\left(  \rho^{\prime}\text{, }\mathcal{R}\right)
\geq0\text{.} \label{5}%
\end{equation}
Finally, combining Eqs. (\ref{2}) and (\ref{5}), we get%
\begin{equation}
\Delta\mathcal{S}_{tot}\overset{\text{def}}{=}\Delta\mathcal{S+H}\left(
p_{k}\right)  \geq0\text{,}%
\end{equation}
that is, exact-QEC does not violate the Second Law because the reduction in
the system's entropy ($\Delta\mathcal{S}<$\ $0$) occurs at the expense of an
increase in the entropy of the environment ($\mathcal{H}\left(  p_{k}\right)
\geq0$) due to the erasure of the demon's measurement record (Landauer's
erasure principle).

We have explained that during a QEC cycle, ancilla-qubits are introduced and
used to record the error syndrome. This can be regarded as a refrigeration
process where entropy which has been introduced into the data-qubits by the
noise gets pumped out into the ancilla-qubits, cooling down the data-qubits.
We have failed to underline that this procedure works provided that the
ancilla-qubits used are cold themselves, otherwise they cannot absorb the
extra entropy from the data. However, ancilla-qubits created at the beginning
of the computation are themselves subject to the noise process. Thus, they
could heat up over time and become worthless for the cooling process. At first
sight, this may appear a big problem. Fortunately, it was recently argued that
a quantum computation needs not necessarily fresh ancilla-qubits supplied from
the outside \cite{ben}. Specifically, it was shown that in the presence of
depolarizing, dephasing and amplitude damping noise models, quantum
computations (without adding fresh ancilla-qubits) are possible for
logarithmic, polynomial and exponential times in the number of available
qubits, respectively.

Returning to our analysis, we may wonder: how does the thermodynamics of a QEC
cycle change when, for instance, the observation (measurement) is not perfect
and the information gain is sub-optimal? How does imperfect discrimination of
non-orthogonal quantum states affect the entropic analysis of a QEC cycle?
These questions will be considered in Section V. In the next section, instead,
we briefly explain the meaning of exact and approximate-QEC schemes.

\section{Exact and approximate quantum error correction}

Formally, there does not exist any QEC scheme that can correct all errors
\cite{hwang}. In other words, only some subsets of all possible errors can be
corrected with a QEC procedure. Therefore, the strategy is to choose certain
subclasses of errors that constitute dominant parts as to-be-corrected ones,
while other classes of errors that constitute negligible parts as
not-to-be-corrected ones. We name exact-QEC an error correction scheme where
it exists a nonzero nontrivial correctable error set which exactly fulfills
the Knill-Laflamme (KL) conditions \cite{nielsen2}. This means that the errors
to be corrected by means of a nondegenerate Pauli basis code have to map the
codeword space to orthogonal spaces if the syndrome is to be detected
unambiguously. The deep reason for this lies in \textit{the process of quantum
measurement which cannot perfectly discriminate among nonorthogonal quantum
states}. However, exact processes exist only as abstract mathematical
concepts. In all practical implementations, the experimenter can only rely
upon some confidence level. Furthermore, there are realistic processes, such
as amplitude damping, where the KL conditions are only approximately satisfied
\cite{nielsen2}. In approximate-QEC, there is no perfectly correctable set of
errors. Only to a certain order (introducing an order of the perturbation
parameter such as the amplitude damping probability parameter $\gamma$), an
error set may satisfy the KL conditions. Interestingly, here the order plays
the role of separating the \textit{correctable} from{} the
\textit{non-correctable} sets (e.g. the Leung et \textit{al}. four-qubit code
\cite{leung}). Thus, the parameter $\gamma$ in approximate-QEC plays the same
role as those usual parameters that separate correctable and non-correctable
sets in exact-QEC, such as the error probability $p$ of a $1$-qubit error. The
main idea in approximate-QEC is to aim at a less than one fidelity
$\mathcal{F}$,%
\begin{equation}
1-\mathcal{F}\leq\mathcal{O}\left(  \epsilon^{1+t}\right)  \text{,}
\label{ccc}%
\end{equation}
where $\epsilon$ denotes a single-qubit error probability (and, $t\geq0$) and
require that the set of approximately reversible error operators has to
include all errors $A_{k}$ with maximum detection probability,%
\begin{equation}
\underset{\left\vert \psi_{in}\right\rangle \in\mathcal{C}}{\max}%
\text{Tr}\left(  \left\vert \psi_{in}\right\rangle \left\langle \psi
_{in}\right\vert A_{k}^{\dagger}A_{k}\right)  \approx\mathcal{O}\left(
\epsilon^{s}\right)  \text{,}%
\end{equation}
with $s\leq t$ and where $\left\vert \psi_{in}\right\rangle $ is a pure input
state and $\mathcal{C}$ is the codespace \cite{leung}. In this scenario, the
exact input state is not recovered but this is indeed not necessary since we
require the achievement of sub-optimal fidelity values solely. Only a good
overlap between the input and the output states is needed. In terms of the
condition on the codespace, it is sufficient that the action of the
recoverable error operators on the codewords lead to approximately mutually
orthogonal quantum states. In particular, if $\lambda_{\max}$ and
$\lambda_{\min}$ are the largest and smallest eigenvalues of $\mathcal{P}%
_{\mathcal{C}}A_{k}^{\dagger}A_{k}\mathcal{P}_{\mathcal{C}}$ considered as an
operator over the codespace $\mathcal{C}$ with projective operator
$\mathcal{P}_{\mathcal{C}}$ ($A_{k}$ are the correctable enlarged error
operators), condition (\ref{ccc}) requires $\lambda_{\max}-$ $\lambda_{\min
}\leq\mathcal{O}\left(  \epsilon^{1+t}\right)  $. Of course, in an exact-QEC
scenario, $\lambda_{\max}-\lambda_{\min}=0$. Such difference is nonzero in the
approximate case due to both the emergence of off-diagonal matrix-terms of
$\mathcal{P}_{\mathcal{C}}A_{k}^{\dagger}A_{k}\mathcal{P}_{\mathcal{C}}$
generated by slight non-orthogonalities and to unequal diagonal terms caused
by imperfect overlaps of correctable errors acting on the different codewords
spanning the codespace $\mathcal{C}$.

How do these slight non-othogonalities and imperfect recovery schemes affect
the entropic analysis of a QEC scheme? In what follows, we partially address
this question together with those raised at the end of Section III.

\section{Quantum error correction and quantum state discrimination}

The reason why perfect discrimination (or, distinguishability) of
non-orthogonal quantum states leads to the violation of the Second Law is that
it would be possible, otherwise, to construct a closed thermodynamic cycle the
sole result of which would be that heat is extracted from an isothermal
reservoir and converted into useful work. For a recent comprehensive
discussion of this violation, we refer to \cite{nori}.

Quantum measurements play a key role within QEC schemes. As a matter of fact,
measurement of quantum systems plays an important role in detecting and
correcting errors in a quantum computation. In particular, when constructing a
quantum error correcting code that can detect and correct a set of errors
$\left\{  A_{k}\right\}  $, we must be able to distinguish the error $A_{a}$
acting on codeword $\left\vert i_{L}\right\rangle $ from the error $A_{b}$
acting on the codeword $\left\vert j_{L}\right\rangle $. Quantum theory does
not allow to unambiguously distinguish non-orthogonal quantum states. Thus,
the erroneous images $A_{a}\left\vert i_{L}\right\rangle $ and $A_{b}%
\left\vert j_{L}\right\rangle $ must be orthogonal if the code is to correctly
distinguish these errors.

The ability to determine the state of a quantum system is not only severely
limited by thermodynamics, as pointed out earlier, but by quantum theory
itself as well. In particular, even if they are drawn from a known set,
non-orthogonal quantum states cannot be discriminated perfectly. The two most
well-known optimum discrimination strategies are the \textit{optimum
unambiguous error-free discrimination strategy} and the \textit{optimum
ambiguous discrimination with minimum error strategy} \cite{hillery, tony}. In
the former procedure, whenever a definitive answer is returned after a
measurement on the state, the result should be unambiguous, at the expense of
allowing inconclusive outcomes to occur. In the latter procedure, instead, one
requires to have conclusive results only. This means that errors are
unavoidable when the states are non-orthogonal. Based on the measurement
outcome, in each single case then a guess has to be made as to what the state
of the quantum system was. This procedure is known as quantum hypothesis
testing\emph{ }\cite{helstrom}. The problem consists in finding the optimum
measurement strategy that minimizes the probability of errors. In general, the
explicit solution to a quantum hypothesis testing, which is an
error-minimizing problem, is not trivial and analytical expressions have been
derived only for a few special cases. For instance, the solution of the
problem of distinguishing two pure non-orthogonal quantum states with minimum
error is considered a pioneering work in quantum detection theory and was
uncovered by Helstrom. The optimal value $\mathcal{P}_{E}\overset{\text{def}%
}{=}\min\mathcal{P}_{\text{err.}}$ of the probability of error $\mathcal{P}%
_{\text{err.}}$ obtained by Helstrom reads \cite{hillery},%
\begin{equation}
\mathcal{P}_{E}\overset{\text{def}}{=}2^{-1}\left[  1-\left(  1-4\eta_{1}%
\eta_{2}\left\vert \left\langle \psi_{1}|\psi_{2}\right\rangle \right\vert
^{2}\right)  ^{\frac{1}{2}}\right]  \text{,} \label{helstrom}%
\end{equation}
where, in general, $\mathcal{P}_{\text{err.}}$ is defined as,%
\begin{equation}
\mathcal{P}_{\text{err.}}\overset{\text{def}}{=}1-\mathcal{P}_{\text{corr.}%
}=1-\sum_{k=1}^{N}\eta_{k}\text{Tr}\left(  \rho_{k}\Pi_{k}\right)  \text{
with, }\sum_{k=1}^{N}\Pi_{k}=I_{D\times D}\text{.}%
\end{equation}
The quantity $D$ denotes the dimensionality of the physical space state,
$\eta_{k}$ are the \textit{a priori} probabilities of occurrence of the
quantum states, $\Pi_{k}$ are the detection operators that characterize the
measurement process and $\rho_{k}$ are the density operators of the $N$ states
of a quantum system. As an illustrative example, consider the following two
pure states $\left\vert \psi_{1}\right\rangle \overset{\text{def}}%
{=}1/2\left\vert 0\right\rangle +\sqrt{3}/2\left\vert 1\right\rangle $ and
$\left\vert \psi_{2}\right\rangle \overset{\text{def}}{=}\sqrt{3}/2\left\vert
0\right\rangle +1/2\left\vert 1\right\rangle $ with equal a priori
probabilities $\eta_{1}=\eta_{2}=1/2$. It turns out that a convenient choice
for the optimal von Neumann measurement operators is given by $\Pi_{1}%
\overset{\text{def}}{=}\left\vert 1\right\rangle \left\langle 1\right\vert $
and $\Pi_{2}\overset{\text{def}}{=}\left\vert 0\right\rangle \left\langle
0\right\vert $ (orthogonal detectors placed symmetrically around $\left\vert
\psi_{1}\right\rangle $ and $\left\vert \psi_{2}\right\rangle $). In this
case, $\mathcal{P}_{E}\overset{\text{def}}{=}\min\mathcal{P}_{\text{err.}%
}=0.25$.

In what follows, being within the ambiguous discrimination strategy framework,
we \textit{perturb} the proof concerning the perfect discrimination of
orthogonal-states in such a way to accommodate \textit{imperfect/approximate}
or, better yet, ambiguous discrimination of non-orthogonal states. We show,
via a simple alternative route, that our reasoning is consistent with standard
arguments that give the square modulus of the overlap of non-orthogonal
quantum states as the essential quantity that limits the effectiveness of
discrimination (with nonvanishing minimum error probability) between quantum
states \cite{dieks}.\ In particular, we check the compatibility of the main
consequence of our analysis with the above-mentioned Helstrom's pioneering
result in the limit of very small probability of error.

\subsection{Discrimination of non-orthogonal states: an old viewpoint
revisited}

Before starting our analysis, let us reconsider the proof establishing that
\textit{it is impossible to unambiguously distinguish non-orthogonal pure
quantum states}. This assertion is proved by contradiction. We assume that
non-orthogonal quantum states can be unambiguously distinguished and show that
this leads to a contradiction. Consider two non-orthogonal states $\left\vert
\psi_{1}\right\rangle $ and $\left\vert \psi_{2}\right\rangle $. Let
$\mathcal{O}$ be an observable represented by the Hermitian operator $\hat{O}$
with eigenvalues $\lambda_{k}$ and projection operators $\Pi_{k}$ such that
its measurement allows to unambiguously distinguish $\left\vert \psi
_{1}\right\rangle $ and $\left\vert \psi_{2}\right\rangle $. This implies that
eigenvalues $\lambda_{\alpha}$ and $\lambda_{\beta}$ exist such that
observation of $\lambda_{\alpha}$ ($\lambda_{\beta}$) unambiguously identifies
$\left\vert \psi_{1}\right\rangle $ ($\left\vert \psi_{2}\right\rangle $) as
the pre-measurement state. Formally, this means that the probability to
observe $\lambda_{\alpha}$ ($\lambda_{\beta}$) when the pre-measurement state
is $\left\vert \psi_{1}\right\rangle $ ($\left\vert \psi_{2}\right\rangle $)
is one,%
\begin{equation}
\left\langle \psi_{1}\left\vert \Pi_{\alpha}\right\vert \psi_{1}\right\rangle
=1\text{ and }\left\langle \psi_{2}\left\vert \Pi_{\beta}\right\vert \psi
_{2}\right\rangle =1\text{,} \label{a0}%
\end{equation}
and thus the probability to observe $\lambda_{\beta}$ ($\lambda_{\alpha}$)
when the pre-measurement state is $\left\vert \psi_{1}\right\rangle $
($\left\vert \psi_{2}\right\rangle $) is zero,%
\begin{equation}
\left\langle \psi_{1}\left\vert \Pi_{\beta}\right\vert \psi_{1}\right\rangle
=0\text{ and }\left\langle \psi_{2}\left\vert \Pi_{\alpha}\right\vert \psi
_{2}\right\rangle =0. \label{a11}%
\end{equation}
Since $\left\vert \psi_{1}\right\rangle $ and $\left\vert \psi_{2}%
\right\rangle $ are assumed to be non-orthogonal, we can write%
\begin{equation}
\left\vert \psi_{2}\right\rangle \overset{\text{def}}{=}c_{1}\left\vert
\psi_{1}\right\rangle +c_{d}\left\vert \psi_{d}\right\rangle \text{,}
\label{a2}%
\end{equation}
where $\left\vert c_{1}\right\vert ^{2}+\left\vert c_{d}\right\vert ^{2}=1$
and $\left\vert \psi_{d}\right\rangle $ is orthogonal to $\left\vert \psi
_{1}\right\rangle $. Observe that $\left\langle \psi_{1}\left\vert \Pi_{\beta
}\right\vert \psi_{1}\right\rangle =0$ implies $\Pi_{\beta}\left\vert \psi
_{1}\right\rangle =0$ since
\begin{equation}
0=\left\langle \psi_{1}\left\vert \Pi_{\beta}\right\vert \psi_{1}\right\rangle
=\left\langle \psi_{1}\left\vert \Pi_{\beta}\Pi_{\beta}\right\vert \psi
_{1}\right\rangle =\left\Vert \Pi_{\beta}\left\vert \psi_{1}\right\rangle
\right\Vert ^{2}\text{,} \label{A}%
\end{equation}
and the only state with zero norm is the null state. Combining (\ref{a2}) with
(\ref{A}) allows us to explicitly evaluate $\left\langle \psi_{2}\left\vert
\Pi_{\beta}\right\vert \psi_{2}\right\rangle $,%
\begin{equation}
\left\langle \psi_{2}\left\vert \Pi_{\beta}\right\vert \psi_{2}\right\rangle
=\left\vert c_{d}\right\vert ^{2}\left\langle \psi_{d}\left\vert \Pi_{\beta
}\right\vert \psi_{d}\right\rangle \text{.} \label{a3}%
\end{equation}
Observe that,%
\begin{equation}
1=\left\langle \psi_{d}\left\vert \psi_{d}\right.  \right\rangle =\left\langle
\psi_{d}\left\vert I\right\vert \psi_{d}\right\rangle =\sum_{k}\left\langle
\psi_{d}\left\vert \Pi_{k}\right\vert \psi_{d}\right\rangle \geq\left\langle
\psi_{d}\left\vert \Pi_{\beta}\right\vert \psi_{d}\right\rangle \text{,}
\label{a4}%
\end{equation}
where the inequality appears since all terms in the sum are non-negative
($\left\langle \psi_{d}\left\vert \Pi_{k}\right\vert \psi_{d}\right\rangle $
is the probability that $\lambda_{k}$ is the measurement outcome when the
pre-measurement state is $\left\vert \psi_{d}\right\rangle $). Combining
(\ref{a3})\ and (\ref{a4}) yields,%
\begin{equation}
1\equiv\left\langle \psi_{2}\left\vert \Pi_{\beta}\right\vert \psi
_{2}\right\rangle =\left\vert c_{d}\right\vert ^{2}\left\langle \psi
_{d}\left\vert \Pi_{\beta}\right\vert \psi_{d}\right\rangle \leq\left\vert
c_{d}\right\vert ^{2}\text{,}%
\end{equation}
that is, $\left\vert c_{d}\right\vert ^{2}\geq1$. However, recall that
$\left\vert c_{1}\right\vert ^{2}+\left\vert c_{d}\right\vert ^{2}=1$.
Therefore, it must be $c_{d}=1$ and $c_{1}=0$, that is $\left\vert \psi
_{2}\right\rangle =\left\vert \psi_{d}\right\rangle $. Thus, for $\left\vert
\psi_{2}\right\rangle $ to be unambiguously distinguishable from $\left\vert
\psi_{1}\right\rangle $, the two pure quantum states must be orthogonal (in
particular, any two orthogonal entangled quantum states can be distinguished
just as well using local operations and classical communication as they can
globally \cite{hardy}). However, we assumed that these states were
non-orthogonal so that we arrived at a contradiction that proves the assertion.

We emphasize that the line of reasoning just presented exhibits two main
features. First, no inconclusive outcome was allowed since the sum of the
measurement operators add up to the unit operator. Second, no ambiguity
(imperfect discrimination) was permitted as evident from Eqs. (\ref{a0}) and
(\ref{a11}).

\subsection{Discrimination of non-orthogonal states: a novel viewpoint}

In what follows, we perturb the above-reconsidered analysis preserving the
first feature but relaxing the second one by introducing an
\textit{ambiguity-factor} $\delta$ which can be ultimately regarded as the
generator of a non-vanishing probability of error within the scheme of optimum
ambiguous discrimination. Specifically, our main working hypothesis is that
Eqs. (\ref{a0}) and (\ref{a11}) assume the following new forms,%
\begin{equation}
\left\langle \psi_{1}\left\vert \Pi_{\alpha}\right\vert \psi_{1}\right\rangle
=1-\delta\text{ and }\left\langle \psi_{2}\left\vert \Pi_{\beta}\right\vert
\psi_{2}\right\rangle =1-\delta\text{,} \label{a6}%
\end{equation}
and thus the probability to observe $\lambda_{\beta}$ ($\lambda_{\alpha}$)
when the pre-measurement state is $\left\vert \psi_{1}\right\rangle $
($\left\vert \psi_{2}\right\rangle $) assumes a non-vanishing value $\delta$,%
\begin{equation}
\left\langle \psi_{1}\left\vert \Pi_{\beta}\right\vert \psi_{1}\right\rangle
=\delta\text{ and }\left\langle \psi_{2}\left\vert \Pi_{\alpha}\right\vert
\psi_{2}\right\rangle =\delta\text{.} \label{a8}%
\end{equation}
The non-orthogonality between $\left\vert \psi_{1}\right\rangle $ and
$\left\vert \psi_{2}\right\rangle $ allows us to reconsider the decomposition
given in Eq. (\ref{a2}). Inserting (\ref{a2}) into the second relation in
(\ref{a6}), we get%
\begin{equation}
1-\delta\overset{\text{def}}{=}\left\langle \psi_{2}\left\vert \Pi_{\beta
}\right\vert \psi_{2}\right\rangle =c_{1}^{\ast}\left\langle \psi
_{1}\left\vert \Pi_{\beta}\right\vert \psi_{2}\right\rangle +c_{1}\left\langle
\psi_{2}\left\vert \Pi_{\beta}\right\vert \psi_{1}\right\rangle -\left\vert
c_{1}\right\vert ^{2}\left\langle \psi_{1}\left\vert \Pi_{\beta}\right\vert
\psi_{1}\right\rangle +\left\vert c_{d}\right\vert ^{2}\left\langle \psi
_{d}\left\vert \Pi_{\beta}\right\vert \psi_{d}\right\rangle \text{.}
\label{a7}%
\end{equation}
For the sake of clarity, we assume that the quantum states considered are
\emph{real}-valued and using Eqs. (\ref{a6}) and (\ref{a8}), Eq.
(\ref{a7})\ becomes%
\begin{equation}
1-\delta=2c_{1}\sqrt{\delta\left(  1-\delta\right)  }-c_{1}^{2}\delta+\left(
1-c_{1}^{2}\right)  \left\langle \psi_{d}\left\vert \Pi_{\beta}\right\vert
\psi_{d}\right\rangle \text{.}%
\end{equation}
However, Eq. (\ref{a4}) implies that $\left\langle \psi_{d}\left\vert
\Pi_{\beta}\right\vert \psi_{d}\right\rangle \leq1$ and, thus, we arrive at
the following inequality constraint relating the \textit{ambiguity-factor}
$\delta$ and the overlap $c_{1}\overset{\text{def}}{=}\left\langle \psi
_{1}\left\vert \psi_{2}\right.  \right\rangle $,%
\begin{equation}
\left(  1+\delta\right)  c_{1}^{2}-2\sqrt{\delta\left(  1-\delta\right)
}c_{1}-\delta\leq0\text{.} \label{ic}%
\end{equation}
Finally, in the limiting case of interest, $\delta\ll1$, the inequality
constraint (\ref{ic}) requires that%
\begin{equation}
\delta\gtrsim\mathcal{\tilde{\eta}}_{12}\left\vert \left\langle \psi
_{1}\left\vert \psi_{2}\right.  \right\rangle \right\vert ^{2}\text{,}
\label{our}%
\end{equation}
with $0\leq\mathcal{\tilde{\eta}}_{12}\overset{\text{def}}{=}\left(
1+\sqrt{2}\right)  ^{-2}\leq1$. Our analysis leads to the conclusion that
$\delta_{\text{min.}}\propto$ $\left\vert \left\langle \psi_{1}\left\vert
\psi_{2}\right.  \right\rangle \right\vert ^{2}$ and confirms that the\textbf{
}square modulus of the overlap of non-orthogonal quantum states is the
essential quantity that limits the effectiveness of discrimination between
quantum states when no inconclusive measurement outcome is permitted
\cite{dieks} (ambiguous QSD, conclusive classification with errors). Observe
that $\left\vert \left\langle \psi_{1}\left\vert \psi_{2}\right.
\right\rangle \right\vert ^{2}\overset{\text{def}}{=}$Tr$\left(  \rho_{1}%
\rho_{2}\right)  $ where $\rho_{k}\overset{\text{def}}{=}\left\vert \psi
_{k}\right\rangle \left\langle \psi_{k}\right\vert $ with $k=1$, $2$ are pure
quantum states ($\rho_{k}^{2}=\rho_{k}$). For the sake of completeness, we
also remark that there are cases where the effectiveness of discrimination
between two non-orthogonal quantum states $\left\vert \psi_{1}\right\rangle $
and $\left\vert \psi_{2}\right\rangle $ is limited by $\left\vert \left\langle
\psi_{1}|\psi_{2}\right\rangle \right\vert $. This happens in a classification
without errors where the modulus of the overlap sets the bound (the so-called
Ivanovich-Dieks-Peres bound, \cite{hillery}). As already pointed out in the
beginning of this section, in this alternative case, the discrimination
procedure enables to infer with certainty whether the system was in the state
$\left\vert \psi_{1}\right\rangle $ or $\left\vert \psi_{2}\right\rangle $ and
leaves a minimum number of cases undecided (unambiguous QSD, inconclusive
classification without errors). As an additional consistency check of our
analysis, we notice that in analogy to the working condition $\delta\ll1$,
setting $\mathcal{P}_{E}\ll1$, and re-arranging Helstrom's formula in
(\ref{helstrom}) together with neglecting higher order infinitesimal terms in
the Taylor-expansion of $\left\vert \left\langle \psi_{1}\left\vert \psi
_{2}\right.  \right\rangle \right\vert ^{2}$ in (\ref{helstrom}), we arrive at%
\begin{equation}
\mathcal{P}_{\text{err.}}\geq\mathcal{P}_{E}\simeq\eta_{1}\eta_{2}\left\vert
\left\langle \psi_{1}\left\vert \psi_{2}\right.  \right\rangle \right\vert
^{2}\text{.} \label{theirs}%
\end{equation}
Upon the reasonable identification of $\delta$ with $\mathcal{P}_{\text{err.}%
}$ (after all, \textit{ambiguity does cause errors}) and for a convenient
choice of measurement operators, we find out that our positive numerical
proportionality factor in (\ref{our}) is less than unity and is compatible
with a suitable choice of a pair of\textit{\ a priori} probabilities $\eta
_{1}$ and $\eta_{2}$ in (\ref{theirs}) (namely, $\eta_{1}=0.78005$ and
$\eta_{2}=0.21995$).

How are these considerations related to the questions asked at the end of
Sections III and IV? An illustrative example may render the idea. Following
\cite{vedral}, assume that only two errors $A_{k}$ with $k=0$, $1$ need to be
corrected and that the imperfect measurement is characterized by two
non-orthogonal quantum states of the apparatus given by $\left\vert
m_{1}\right\rangle \overset{\text{def}}{=}\left\vert 0\right\rangle $ and
$\left\vert m_{2}\right\rangle \overset{\text{def}}{=}\xi\sqrt{\delta
}\left\vert 0\right\rangle +\sqrt{1-\xi^{2}\delta}\left\vert 1\right\rangle $
where $\left\langle m_{1}\left\vert m_{2}\right.  \right\rangle =$ $\xi
\sqrt{\delta}$ with $0\leq\delta\leq1$ and $0\leq\xi\leq\delta^{-\frac{1}{2}}%
$. The negativity of the rate of change of the von Neumann erasure entropy
with respect to the quantum overlap $\left\langle m_{1}\left\vert
m_{2}\right.  \right\rangle $,%
\begin{equation}
\frac{\partial\mathcal{S}\left(  \frac{\left\vert m_{1}\right\rangle
\left\langle m_{1}\right\vert +\left\vert m_{2}\right\rangle \left\langle
m_{2}\right\vert }{\text{Tr}\left(  \left\vert m_{1}\right\rangle \left\langle
m_{1}\right\vert +\left\vert m_{2}\right\rangle \left\langle m_{2}\right\vert
\right)  }\right)  }{\partial\left(  \left\langle m_{1}\left\vert
m_{2}\right.  \right\rangle \right)  }\propto\left\langle m_{1}\left\vert
m_{2}\right.  \right\rangle \log\left(  \frac{1-\left\langle m_{1}\left\vert
m_{2}\right.  \right\rangle }{1+\left\langle m_{1}\left\vert m_{2}\right.
\right\rangle }\right)  \leq0\text{,} \label{sub}%
\end{equation}
leads to the conclusion that the bigger is the quantum overlap of
non-orthogonal quantum apparatus states, the smaller is the erasure entropy
and, because of Landauer's principle, the smaller is the quantum information
gain. Imperfect discrimination (see Eq. (\ref{our})) leads to sub-optimal
quantum information gain (see Eq. (\ref{sub})) which is a fingerprint of
approximate-QEC (see Eq. (\ref{ccc})). More generally, we might state that in
terms of the quantum discrimination formalism, exact-QEC may be regarded as
error-free and conclusive (syndrome extraction) for the set of correctable
errors while it appears inconclusive for the discrimination between the sets
of correctable and non-correctable errors. On the other hand, approximate-QEC
is conclusive with a small finite probability of error for the set of
correctable errors. Instead, a large probability of error occurs and no
conclusive discrimination happens between the sets of correctable and
non-correctable errors, just as in the case of exact-QEC.

For the sake of completeness, we finally remark that we have limited our
considerations to deterministic QEC schemes. However, probabilistic QEC
schemes could have been considered as well \cite{ueda}. In such schemes,
characterized by probabilistically reversible measurements, quantum codes may
correct errors in such a manner that the overall probability of success is
less than one.

\section{Concluding Comments}

In this work, we discussed the relevance of entropy, information and the
Second Law of thermodynamics in a QEC cycle regarded as a special type of a
Maxwell's demon. Our main effort was focused at clarifying the role played by
the process of quantum measurement (which cannot perfectly discriminate among
non-orthogonal states) in the entropic analysis of an approximate-QEC cycle.
We have provided semi-quantitative reasoning for explaining the reason why the
square modulus of the overlap of non-orthogonal quantum states is the
essential quantity that limits the effectiveness of discrimination between
quantum states when no inconclusive measurement outcome is permitted. Finally,
using this point, we have stressed the link among perfect (imperfect)
discrimination, optimal (sub-optimal) quantum information gain and exact-
(approximate-) QEC.

We hope to deepen our formal understanding and strengthen our quantitative
analysis concerning these information-theoretic links in forthcoming efforts.
Specifically, we would like to recast both exact and approximate-QEC schemes
in a state discrimination formalism for (stabilizer) mixed quantum states
represented by density operators (since the entities to be discriminated in
QEC are actually subspaces rather than pure quantum states) and, hopefully,
quantify analogies and differences between the two schemes in quantum
(relative) entropic terms. For the time being, we limit ourself to provide a
simple illustrative example that exhibits the link between QEC and quantum
state discrimination of orthogonal stabilizer mixed states in the appendix.

We would like to conclude with an illuminating reply of von Neumann to the
canonical question about the impossibility of constructing thinking machines
as reported by Jaynes in \cite{jay-science}: \textit{If you will tell me
precisely what it is that a machine cannot do, then I can always make a
machine which will do just that}. While von Neumann's remark may be
unquestionable on purely conceptual grounds, the actual realization of a less
demanding (and, perhaps, more useful) non-thinking \textit{quantum machine} is
turning out to be a highly nontrivial task to achieve \cite{ric}. In any case,
the lesson here seems to be that such quantum machines together with all its
embodied computational schemes must function obeying accepted physical laws
just as it happens with QEC and the Second Law. Ultimately, what seems
undisputable is that we cannot exit the realm of accepted laws of physics, or
rules of inductive inference, as someone may argue.

\begin{acknowledgments}
We acknowledge that a very preliminary version of this work was originally
presented at MaxEnt 2012, the 32nd International Workshop on Bayesian
Inference and Maximum Entropy Methods in Science and Engineering held at the
Max-Planck-Institut fur Plasmaphysik (IPP) in Garching bei Munchen, Germany.
CC thanks Hussain Zaidi for calling his attention to reference \cite{hillery}.
We thank the ERA-Net CHIST-ERA project HIPERCOM for financial support.
\end{acknowledgments}

\appendix

\section{An illustrative example}

In both exact- and approximate-QEC, the entities to be discriminated are
subspaces rather than states. For this reason, in order to properly recast the
properties of QEC in terms of quantum state discrimination (QSD), general
mixed states should be taken into consideration. For quantum stabilizer codes,
such subspaces are the so-called stabilizer mixed states. These states have
zero overlap for exact-QEC and correctable errors while they can exhibit
non-zero overlap for approximate-QEC and correctable (recoverable) errors. For
the discrimination of orthogonal subspaces, we follow \cite{sb}; for the
discrimination of non-orthogonal subspaces, we refer to \cite{jab}. For a
geometric approach to QSD, we refer to \cite{damian}. We remark that when
passing from pure to mixed states quantum discrimination, the realm of
possible scenarios to consider becomes more complex and additional care is
needed. For instance, it is possible to construct mixed states which cannot be
distinguished perfectly locally, despite being orthogonal \cite{prl}.
Moreover, quantum mixed states cannot be unambiguously discriminated
\cite{jarom}, in general. That said, it is not our intention to provide here a
complete description that concerns the recasting of the general QEC problem
into the general QSD one. However, in what follows, we shall present a simple
illustrative example where stabilizer mixed states are employed and a link
between exact-QEC and QSD of orthogonal subspaces is made transparent.

\subsection{Exact-QEC and QSD}

For the sake of reasoning, we restrict our considerations to the bit-flip (or,
equivalently, phase-flip/dephasing) noise model and to the three-qubit
bit-flip repetition code \cite{nielsen2}. \textbf{ }We are aware that a pure
dephasing channel, with no other sources of noise at all, is physically
improbable. However, in many physical systems, dephasing is indeed the
dominant error source \cite{ben}.\textbf{ }The operator sum representation of
the enlarged quantum channel after encoding reads,%
\begin{equation}
\Lambda_{\text{bit-flip}}\left(  \rho\right)  \overset{\text{def}}{=}%
\sum_{k=0}^{7}A_{k}\rho A_{k}^{\dagger}\text{,}%
\end{equation}
where the eight enlarged error operators are given by,%
\begin{align}
&  A_{0}\overset{\text{def}}{=}\sqrt{\left(  1-p\right)  ^{3}}I^{1}\otimes
I^{2}\otimes I^{3}\text{, }A_{1}\overset{\text{def}}{=}\sqrt{p\left(
1-p\right)  ^{2}}X^{1}\otimes I^{2}\otimes I^{3}\text{, }A_{2}\overset
{\text{def}}{=}\sqrt{p\left(  1-p\right)  ^{2}}I^{1}\otimes X^{2}\otimes
I^{3}\text{,}\nonumber\\
& \nonumber\\
&  A_{3}\overset{\text{def}}{=}\sqrt{p\left(  1-p\right)  ^{2}}I^{1}\otimes
I^{2}\otimes X^{3}\text{, }A_{4}\overset{\text{def}}{=}\sqrt{p^{2}\left(
1-p\right)  }X^{1}\otimes X^{2}\otimes I^{3}\text{, }A_{5}\overset{\text{def}%
}{=}\sqrt{p^{2}\left(  1-p\right)  }X^{1}\otimes I^{2}\otimes X^{3}%
\text{,}\nonumber\\
& \nonumber\\
&  A_{6}\overset{\text{def}}{=}\sqrt{p^{2}\left(  1-p\right)  }I^{1}\otimes
X^{2}\otimes X^{3}\text{, }A_{7}\overset{\text{def}}{=}\sqrt{p^{3}}%
X^{1}\otimes X^{2}\otimes X^{3}\text{.}%
\end{align}
The three-qubit bit-flip repetition code is characterized by a two-dimensional
\emph{complex} subspace of the eight-dimensional Hilbert space $\mathcal{H}%
_{2}^{3}$ and is spanned by the codewords,%
\begin{equation}
\left\vert 0_{L}\right\rangle \overset{\text{def}}{=}\left\vert
000\right\rangle \text{ and, }\left\vert 1_{L}\right\rangle \overset
{\text{def}}{=}\left\vert 111\right\rangle \text{.}%
\end{equation}
This code is capable of error-correcting the following four enlarged errors,%
\begin{equation}
\mathcal{A}_{\text{correctable}}=\left\{  A_{0}\text{, }A_{1}\text{, }%
A_{2}\text{, }A_{3}\right\}  \text{.}%
\end{equation}
For any error operator in $\mathcal{A}_{\text{correctable}}$, the standard
Knill-Laflamme error correction conditions are exactly fulfilled \cite{kl},%
\begin{equation}
\left\langle i_{L}|A_{k}^{\dagger}A_{m}|j_{L}\right\rangle =\delta_{ij}%
\alpha_{km}\text{,}%
\end{equation}
where $\alpha_{km}$ are the components of a density operator, $i$,
$j\in\left\{  0\text{, }1\right\}  $ and $k$, $m\in\left\{  0\text{, }1\text{,
}2\text{, }3\right\}  $. The Hilbert space $\mathcal{H}_{2}^{3}$ can be
decomposed as the direct sum of two orthogonal four-dimensional \emph{complex}
Hilbert subspaces (or, equivalently, four two-dimensional \emph{complex}
Hilbert subspaces),%
\begin{align}
\mathcal{H}_{2}^{3}  &  =V^{0_{L}}\oplus V^{1_{L}}\nonumber\\
& \nonumber\\
&  =\text{Span}\left\{  A_{k}\left\vert 0_{L}\right\rangle \right\}
\oplus\text{Span}\left\{  A_{k}\left\vert 1_{L}\right\rangle \right\}
\nonumber\\
& \nonumber\\
&  =\text{Span}\left\{  A_{0}\left\vert 0_{L}\right\rangle \text{, }%
A_{1}\left\vert 0_{L}\right\rangle \text{, }A_{2}\left\vert 0_{L}\right\rangle
\text{, }A_{3}\left\vert 0_{L}\right\rangle \right\}  \oplus\text{Span}%
\left\{  A_{0}\left\vert 1_{L}\right\rangle \text{, }A_{1}\left\vert
1_{L}\right\rangle \text{, }A_{2}\left\vert 1_{L}\right\rangle \text{, }%
A_{3}\left\vert 1_{L}\right\rangle \right\} \nonumber\\
& \nonumber\\
&  =\text{Span}\left\{  A_{0}\left\vert 0_{L}\right\rangle \text{, }%
A_{0}\left\vert 1_{L}\right\rangle \right\}  \oplus\text{Span}\left\{
A_{1}\left\vert 0_{L}\right\rangle \text{, }A_{1}\left\vert 1_{L}\right\rangle
\right\}  \oplus\text{Span}\left\{  A_{2}\left\vert 0_{L}\right\rangle \text{,
}A_{2}\left\vert 1_{L}\right\rangle \right\}  \oplus\text{Span}\left\{
A_{3}\left\vert 0_{L}\right\rangle \text{, }A_{3}\left\vert 1_{L}\right\rangle
\right\} \nonumber\\
& \nonumber\\
&  =\mathcal{S}_{A_{0}}\oplus\mathcal{S}_{A_{1}}\oplus\mathcal{S}_{A_{2}%
}\oplus\mathcal{S}_{A_{3}}\nonumber\\
&  =%
{\textstyle\bigoplus\limits_{k=0}^{3}}
\mathcal{S}_{A_{k}}\text{,}%
\end{align}
with $A_{k}\in\mathcal{A}_{\text{correctable}}$ and where the orthogonal
subspaces $\mathcal{S}_{A_{k}}$ are the two-dimensional \emph{complex}
subspaces of $\mathcal{H}_{2}^{3}$ defined as,%
\begin{equation}
\mathcal{S}_{A_{k}}\overset{\text{def}}{=}\text{Span}\left\{  A_{k}\left\vert
0_{L}\right\rangle \text{, }A_{k}\left\vert 1_{L}\right\rangle \right\}
\text{.} \label{sak}%
\end{equation}
The standard four QEC recovery operators are given by,%
\begin{align}
R_{0}  &  \equiv R_{A_{0}}\overset{\text{def}}{=}\left\vert 0_{L}\right\rangle
\left\langle 0_{L}\right\vert +\left\vert 1_{L}\right\rangle \left\langle
1_{L}\right\vert \text{, }R_{1}\equiv R_{A_{1}}\overset{\text{def}}%
{=}\left\vert 0_{L}\right\rangle \left\langle 100\right\vert +\left\vert
1_{L}\right\rangle \left\langle 011\right\vert \text{,}\nonumber\\
& \nonumber\\
R_{2}  &  \equiv R_{A_{2}}\overset{\text{def}}{=}\left\vert 0_{L}\right\rangle
\left\langle 010\right\vert +\left\vert 1_{L}\right\rangle \left\langle
101\right\vert \text{, }R_{3}\equiv R_{A_{3}}\overset{\text{def}}{=}\left\vert
0_{L}\right\rangle \left\langle 001\right\vert +\left\vert 1_{L}\right\rangle
\left\langle 110\right\vert \text{,} \label{recov}%
\end{align}
where,%
\begin{equation}
\sum_{k=0}^{3}R_{k}^{\dagger}R_{k}=\mathcal{I}_{8\times8}\text{.}%
\end{equation}
We observe that the stabilizer mixed quantum states associated with the
bit-flip noise model when the error correction is performed by means of the
three-qubit bit-flip repetition code are given by,%
\begin{equation}
\rho_{j}\equiv\rho_{A_{j}}\overset{\text{def}}{=}\frac{A_{j}\mathcal{P}%
_{\mathcal{C}}A_{j}^{\dagger}}{2}\text{,} \label{a10}%
\end{equation}
with $j\in\left\{  0\text{,..., }7\right\}  $ and where $\mathcal{P}%
_{\mathcal{C}}\overset{\text{def}}{=}\left\vert 0_{L}\right\rangle
\left\langle 0_{L}\right\vert +\left\vert 1_{L}\right\rangle \left\langle
1_{L}\right\vert $ is the projector on the codespace. To be explicit, we have%
\begin{align}
&  \rho_{0}\overset{\text{def}}{=}\frac{\left\vert 000\right\rangle
\left\langle 000\right\vert +\left\vert 111\right\rangle \left\langle
111\right\vert }{2}\text{, }\rho_{1}\overset{\text{def}}{=}\frac{\left\vert
100\right\rangle \left\langle 100\right\vert +\left\vert 011\right\rangle
\left\langle 011\right\vert }{2}\text{, }\rho_{2}\overset{\text{def}}{=}%
\frac{\left\vert 010\right\rangle \left\langle 010\right\vert +\left\vert
101\right\rangle \left\langle 101\right\vert }{2}\text{,}\nonumber\\
& \nonumber\\
&  \rho_{3}\overset{\text{def}}{=}\frac{\left\vert 001\right\rangle
\left\langle 001\right\vert +\left\vert 110\right\rangle \left\langle
110\right\vert }{2}\text{, }\rho_{4}\overset{\text{def}}{=}\frac{\left\vert
110\right\rangle \left\langle 110\right\vert +\left\vert 001\right\rangle
\left\langle 001\right\vert }{2}\text{, }\rho_{5}\overset{\text{def}}{=}%
\frac{\left\vert 101\right\rangle \left\langle 101\right\vert +\left\vert
010\right\rangle \left\langle 010\right\vert }{2}\text{,}\nonumber\\
& \nonumber\\
&  \rho_{6}\overset{\text{def}}{=}\frac{\left\vert 011\right\rangle
\left\langle 011\right\vert +\left\vert 100\right\rangle \left\langle
100\right\vert }{2}\text{, }\rho_{7}\overset{\text{def}}{=}\frac{\left\vert
111\right\rangle \left\langle 111\right\vert +\left\vert 000\right\rangle
\left\langle 000\right\vert }{2}\text{,} \label{a111}%
\end{align}
and we notice that $\rho_{0}=\rho_{7}$, $\rho_{1}=\rho_{6}$, $\rho_{2}%
=\rho_{5}$, $\rho_{3}=\rho_{4}$. We also stress that the impossibility to
discriminate between $\rho_{A_{j}}$ and $\rho_{A_{j^{\prime}}}$ can be
ascribed to the fact that $\left\{  A_{j}\text{, }A_{j^{\prime}}\right\}  $ is
not a correctable set of two enlarged error operators. For the set of
correctable errors $\left\{  A_{j}\right\}  $ with $j\in\left\{  0\text{,
}1\text{, }2\text{, }3\right\}  $, we obtain%
\begin{equation}
\rho_{j}\rho_{j^{\prime}}=\frac{1}{2}\rho_{j}\delta_{jj^{\prime}}\text{ and,
}\mathcal{O}_{jj^{\prime}}\overset{\text{def}}{=}\text{Tr}\left(  \rho_{j}%
\rho_{j^{\prime}}\right)  =\frac{1}{2}\delta_{jj^{\prime}}\text{.} \label{lim}%
\end{equation}
From Eq. (\ref{lim}), we note that the mixed stabilizer states that correspond
to the set of correctable errors have zero quantum overlap $\mathcal{O}$.

In terms of local discrimination of orthogonal subspaces, it turns out that a
necessary\emph{\ }condition for perfect LOCC (local operations and classical
communication, \cite{nielsen2})\ state discrimination is the following: if the
orthogonal quantum mixed states $\rho_{1}$,..., $\rho_{\bar{k}}$ are perfectly
distinguishable by LOCC then it is necessary that there exists a separable
POVM (positive-operator valued measure, \cite{nielsen2}) $\Pi=\left\{  \Pi
_{1}\text{,..., }\Pi_{\bar{k}}\right\}  $ such that \cite{sb},%
\begin{equation}
\text{Tr}\left(  \Pi_{i}\rho_{j}\right)  =\delta_{ij}\text{, }\forall i\text{,
}j\in\left\{  1\text{,..., }\bar{k}\right\}  \text{.} \label{lim2}%
\end{equation}
We recall that a separable measurement $\Pi\overset{\text{def}}{=}\left\{
\Pi_{1}\text{,..., }\Pi_{\bar{k}}\right\}  $ on a Hilbert space $\mathcal{H}$
is a POVM such that,%
\begin{equation}
\sum_{i=1}^{\bar{k}}\Pi_{i}=\mathcal{I}_{\mathcal{H}}\text{,}%
\end{equation}
where $\Pi_{i}$ is a separable, positive semi-definite operator for every $i$
and $\mathcal{I}_{\mathcal{H}}$ is the identity operator on $\mathcal{H}$. In
our simple illustrative example, we can rewrite the density matrices $\rho
_{k}$ in Eq. (\ref{a10}) as,%
\begin{equation}
\rho_{k}\overset{\text{def}}{=}\frac{1}{\dim_{%
\mathbb{C}
}\left(  \mathcal{S}_{A_{k}}\right)  }\mathcal{P}_{\mathcal{S}_{A_{k}}%
}\text{,}%
\end{equation}
where $\mathcal{P}_{\mathcal{S}_{A_{k}}}$ are the projectors onto the
orthogonal subspaces $\mathcal{S}_{A_{k}}$ in Eq. (\ref{sak}). The operators
$\mathcal{P}_{\mathcal{S}_{A_{k}}}$ are given by,
\begin{equation}
\mathcal{P}_{\mathcal{S}_{A_{k}}}\overset{\text{def}}{=}A_{k}\left\vert
0_{L}\right\rangle \left\langle 0_{L}\right\vert A_{k}^{\dagger}%
+A_{k}\left\vert 1_{L}\right\rangle \left\langle 1_{L}\right\vert
A_{k}^{\dagger}=A_{k}\left(  \left\vert 0_{L}\right\rangle \left\langle
0_{L}\right\vert +\left\vert 1_{L}\right\rangle \left\langle 1_{L}\right\vert
\right)  A_{k}^{\dagger}=A_{k}\mathcal{P}_{\mathcal{C}}A_{k}^{\dagger}\text{.}%
\end{equation}
The set of orthogonal quantum stabilizer mixed states $\mathcal{D}%
\overset{\text{def}}{=}\left\{  \rho_{0}\text{, }\rho_{1}\text{, }\rho
_{2}\text{, }\rho_{3}\right\}  $ with the $\rho_{i}$s defined in Eq.
(\ref{a111}) could be discriminated by the separable measurement $\Pi
\overset{\text{def}}{=}\left\{  \Pi_{0}\text{, }\Pi_{1}\text{, }\Pi_{2}\text{,
}\Pi_{3}\right\}  $,%
\begin{equation}
\Pi_{k}\overset{\text{def}}{=}R_{k}^{\dagger}R_{k}\text{ with, }\sum_{k=0}%
^{3}\Pi_{k}=\mathcal{I}_{8\times8}\text{.}%
\end{equation}
To be explicit, we have%
\begin{align}
\Pi_{0}  &  =R_{0}^{\dagger}R_{0}=\left\vert 000\right\rangle \left\langle
000\right\vert +\left\vert 111\right\rangle \left\langle 111\right\vert
\text{, }\Pi_{1}=R_{1}^{\dagger}R_{1}=\left\vert 100\right\rangle \left\langle
100\right\vert +\left\vert 011\right\rangle \left\langle 011\right\vert
\text{,}\nonumber\\
& \nonumber\\
\Pi_{2}  &  =R_{2}^{\dagger}R_{2}=\left\vert 010\right\rangle \left\langle
010\right\vert +\left\vert 101\right\rangle \left\langle 101\right\vert
\text{, }\Pi_{3}=R_{3}^{\dagger}R_{3}=\left\vert 001\right\rangle \left\langle
001\right\vert +\left\vert 110\right\rangle \left\langle 110\right\vert
\text{,}%
\end{align}
with the recovery operators $\left\{  R_{k}\right\}  $ with $k\in\left\{
0\text{, }1\text{, }2\text{, }3\right\}  $ defined in Eq. (\ref{recov}). A
simple check allows us to conclude that,%
\begin{equation}
\text{Tr}\left(  \Pi_{l}\rho_{m}\right)  =\delta_{lm}\text{, }\forall l\text{,
}m\in\left\{  0\text{, }1\text{, }2\text{, }3\right\}  \text{.}%
\end{equation}
Thus, the necessary conditions for perfect LOCC mixed state discrimination are
satisfied for the set of correctable errors. For further technical details on
perfect local discrimination of orthogonal quantum states we refer to
\cite{sb}.

\subsection{Approximate-QEC and QSD}

It would be interesting to extend these above-presented considerations to
approximate-QEC and mixed state discrimination as well. For instance, we might
consider the amplitude damping (AD) noise model and error correction performed
by means of the four-qubit Leung et \textit{al}. code \cite{leung}. In
particular, it would be worthwhile studying the manner in which Eqs.
(\ref{lim}) and (\ref{lim2}) change in the framework of approximate-QEC.

The AD\ channel is the simplest nonunital channel whose Kraus operators cannot
be described by (unitary) Pauli operations \cite{nielsen2}. The two Kraus
operators for AD noise are given by $A_{0}\overset{\text{def}}{=}%
I-\mathcal{O}\left(  \gamma\right)  $ and $A_{1}\overset{\text{def}}{=}%
\sqrt{\gamma}\left\vert 0\right\rangle \left\langle 1\right\vert $ where
$\gamma$ denotes the damping rate. As we may observe, there is no simple way
of reducing $A_{1}$ to one Pauli error operator since $\left\vert
0\right\rangle \left\langle 1\right\vert $ is not normal. In the case of
amplitude damping, we model the environment as starting in the $\left\vert
0\right\rangle $ state as it were at zero temperature. This quantum noisy
channel is defined as \cite{nielsen2},%
\begin{equation}
\Lambda_{\text{AD}}\left(  \rho\right)  \overset{\text{def}}{=}\sum_{k=0}%
^{1}A_{k}\rho A_{k}^{\dagger}\text{,}%
\end{equation}
where the Kraus error operators $A_{k}$ read,
\begin{equation}
A_{0}\overset{\text{def}}{=}\frac{1}{2}\left[  \left(  1+\sqrt{1-\gamma
}\right)  I+\left(  1-\sqrt{1-\gamma}\right)  \sigma_{z}\right]  \text{ and,
}A_{1}\overset{\text{def}}{=}\frac{\sqrt{\gamma}}{2}\left(  \sigma_{x}%
+i\sigma_{y}\right)  \text{,} \label{a17}%
\end{equation}
respectively. The $\left(  2\times2\right)  $-matrix representation of the
$A_{k}$ operators is given by,%
\begin{equation}
A_{0}=\left(
\begin{array}
[c]{cc}%
1 & 0\\
0 & \sqrt{1-\gamma}%
\end{array}
\right)  \text{ and, }A_{1}=\left(
\begin{array}
[c]{cc}%
0 & \sqrt{\gamma}\\
0 & 0
\end{array}
\right)  \text{.}%
\end{equation}
The action of the $A_{k}$ with $k\in\left\{  0\text{, }1\right\}  $ operators
on the computational basis vectors $\left\vert 0\right\rangle $ and
$\left\vert 1\right\rangle $ reads,%
\begin{equation}
A_{0}\left\vert 0\right\rangle =\left\vert 0\right\rangle \text{, }%
A_{0}\left\vert 1\right\rangle =\sqrt{1-\gamma}\left\vert 1\right\rangle
\text{, }%
\end{equation}
and,%
\begin{equation}
A_{1}\left\vert 0\right\rangle \equiv0\text{, }A_{1}\left\vert 1\right\rangle
=\sqrt{\gamma}\left\vert 0\right\rangle \text{,}%
\end{equation}
respectively. The codewords of the Leung et \textit{al}. $\left[  \left[
4\text{, }1\right]  \right]  $ quantum code are given by \cite{leung},%
\begin{equation}
\left\vert 0_{L}\right\rangle \overset{\text{def}}{=}\frac{1}{\sqrt{2}}\left(
\left\vert 0000\right\rangle +\left\vert 1111\right\rangle \right)  \text{
and, }\left\vert 1_{L}\right\rangle \overset{\text{def}}{=}\frac{1}{\sqrt{2}%
}\left(  \left\vert 0011\right\rangle +\left\vert 1100\right\rangle \right)
\text{.}%
\end{equation}
We notice that in the case of exact-QEC, the first relation in Eq. (\ref{lim})
can be rewritten as%
\begin{equation}
\rho_{j}\rho_{j^{\prime}}=\frac{A_{j}\mathcal{P}_{\mathcal{C}}A_{j}^{\dagger}%
}{2}\frac{A_{j^{\prime}}\mathcal{P}_{\mathcal{C}}A_{j^{\prime}}^{\dagger}}%
{2}=\frac{1}{4}A_{j}\left(  \mathcal{P}_{\mathcal{C}}A_{j}^{\dagger
}A_{j^{\prime}}\mathcal{P}_{\mathcal{C}}\right)  A_{j^{\prime}}^{\dagger
}\text{,}%
\end{equation}
with,%
\begin{equation}
\mathcal{P}_{\mathcal{C}}A_{j}^{\dagger}A_{j^{\prime}}\mathcal{P}%
_{\mathcal{C}}=\alpha_{jj^{\prime}}\mathcal{P}_{\mathcal{C}}\text{.}
\label{pppp}%
\end{equation}
In the approximate-QEC framework, it can be shown that Eq. (\ref{pppp}) can be
replaced by \cite{beny},%
\begin{equation}
\mathcal{P}_{\mathcal{C}}A_{j}^{\dagger}A_{j^{\prime}}\mathcal{P}%
_{\mathcal{C}}=\alpha_{jj^{\prime}}\mathcal{P}_{\mathcal{C}}+\mathcal{P}%
_{\mathcal{C}}\hat{B}_{jj^{\prime}}\mathcal{P}_{\mathcal{C}}=\mathcal{P}%
_{\mathcal{C}}\left(  \alpha_{jj^{\prime}}I+\hat{B}_{jj^{\prime}}\right)
\mathcal{P}_{\mathcal{C}}\text{,} \label{za}%
\end{equation}
where $\alpha_{jj^{\prime}}$ and $B_{jj^{\prime}}$ can be regarded as the
higher and lower order (with respect to the small parameter that parametrizes
the errors that characterize the noise model) components of a density
operator, respectively. For example, consider the correctable enlarged error
operator $\tilde{A}_{0}\overset{\text{def}}{=}A_{0}\otimes A_{0}\otimes
A_{0}\otimes A_{0}$ with $A_{0}$ defined in Eq. (\ref{a17}). After some
algebra, it follows that Eq. (\ref{za}) gives%
\begin{equation}
\mathcal{P}_{\mathcal{C}}\tilde{A}_{0}^{\dagger}\tilde{A}_{0}\mathcal{P}%
_{\mathcal{C}}=\alpha_{00}\mathcal{P}_{\mathcal{C}}+\mathcal{P}_{\mathcal{C}%
}\hat{B}_{00}\mathcal{P}_{\mathcal{C}}\text{,}%
\end{equation}
where $\alpha_{00}\overset{\text{def}}{=}1-2\gamma$ and,%
\begin{equation}
\hat{B}_{00}\overset{\text{def}}{=}\left(  3\gamma^{2}-2\gamma^{3}+\frac{1}%
{2}\gamma^{4}\right)  \left\vert 0_{L}\right\rangle \left\langle
0_{L}\right\vert +\gamma^{2}\left\vert 1_{L}\right\rangle \left\langle
1_{L}\right\vert \text{.}%
\end{equation}
In the exact case, we recall that the operators $\hat{B}_{jj^{\prime}}$
vanish. We also point out that unlike the exact-QEC case where all recoverable
errors lead to orthogonal mixed stabilizer states that could be perfectly
discriminated in principle, in the approximate-case we record the emergence of
nonvanishing quantum overlaps between mixed stabilizer states corresponding to
correctable ($\tilde{A}_{0}$, for instance) and non-correctable enlarged
errors ($\tilde{A}_{15}\overset{\text{def}}{=}A_{1}\otimes A_{1}\otimes
A_{1}\otimes A_{1}$ with $A_{1}$ defined in Eq. (\ref{a17}), for instance).
Thus, even non-correctable errors could be in principle partially recovered in
the approximate-QEC setting. While this non-orthogonality of quantum states
can be advantageous, its handling certainly requires extra-care when employed
in the context of the state discrimination formalism. From these simple
considerations, we are lead to believe that recasting approximate-QEC into the
quantum state discrimination formalism awaits additional thinking.

As stated in our Concluding Comments, it is our intention to provide a
detailed investigation of these issues in forthcoming efforts.


\begin{thebibliography}{99}                                                                                               %


\bibitem {popescu}P. Skrzypczyk, A. J. Short, and S. Popescu,
\emph{Thermodynamics for individual quantum systems}, arXiv:quant-ph/1307.1558 (2013).

\bibitem {ferdy}Fernando G. S. L. Brandao, M. Horodecki, N. H. Y. Ng, J.
Oppenheim, and S. Wehner, \emph{The second laws of quantum thermodynamics},
arXiv:quant-ph/1305.5278 (2013).

\bibitem {renato}M. Tomamichel and R. Renner, \emph{Uncertainty relation for
smooth entropies}, Phys. Rev. Lett. \textbf{106}, 110506 (2011).

\bibitem {caticha}A. Caticha, \emph{Entropic dynamics, time and quantum
theory}, J. Phys. \textbf{A44}, 225303 (2011).

\bibitem {winter}S. Wehner and A. Winter, \emph{Entropic uncertainty
relations}-\emph{A survey}, New J. Phys. \textbf{12}, 025009 (2010).

\bibitem {beretta}G. P. Beretta, E. P. Gyftopoulos, J. L. Park, and G. N.
Hatsopoulos, \emph{Quantum thermodynamics. A new equation of motion for a
single constituent of matter}, Nuovo Cimento \textbf{B82}, 169 (1984).

\bibitem {einstein}A. Einstein, B. Podolsky, and N. Rosen, \emph{Can
quantum-mechanical description of physical reality be considered complete?},
Phys. Rev. \textbf{47}, 777 (1935).

\bibitem {szilard}L. Szilard, \emph{On the decrease of entropy in a
thermodynamic system by the intervention of intelligent beings}, Z. Phys.
\textbf{53}, 840 (1929).

\bibitem {landauer1}R. Landauer,\emph{ Irreversibility and heat generation in
the computing process}, IBM J. Res. Dev. \textbf{5, }183 (1961).

\bibitem {bennett}C. H. Bennett, \emph{Notes on Landauer's principle,
reversible computation, and Maxwell's demon}, \textit{Studies in History and
Philosophy of Modern Physics} \textbf{34}, 501 (2003).

\bibitem {zurek}W. H. Zurek, \emph{Maxwell's demon, Slizard's engine and
quantum measurements}, in G. T. Moore and M. O. Scully, Frontiers of
Non-equilibrium Statistical Physics (Plenum Press, New York, 1984), pp. 151-161.

\bibitem {charlie}C. H. Bennett, \emph{Quantum information: qubits and quantum
error correction}, Int. J. Theor. Phys. \textbf{42}, 153 (2003).

\bibitem {vedral}V. Vedral, \emph{Landauer's principle, error correction and
entanglement}, \textit{Proc. R. Soc. Lond.} \textbf{A456}, 969 (2000).

\bibitem {jauch}J. M. Jauch and J. B. Baron, \emph{Entropy, information and
Szilard's paradox}, Helv. Phys. Acta \textbf{45}, 220 (1972).

\bibitem {berger}J. Berger, \emph{Szilard's demon revisited}, Int. J. Theor.
Phys. \textbf{29}, 985 (1990).

\bibitem {goto}E. Goto, W. Hioe, and M. Hosoya,\emph{ Physical limits to
quantum flux parametron operation}, Physica \textbf{C185-189}, 385 (1991).

\bibitem {tribus}O. Costa de Beauregard and M. Tribus, \emph{Information
theory and thermodynamics}, Helv. Phys. Acta \textbf{47}, 238 (1974).

\bibitem {peres1}A. Peres, \emph{Thermodynamic constraints on quantum axioms},
in \textit{Complexity, Entropy and the Physics of Information} ed. W. H.
Zurek, pp. 345-355 (1991).

\bibitem {peres2}A. Peres, \emph{Quantum Theory: Concepts and Methods}, Kluver
Academic Publishers (1995).

\bibitem {john}J. von Neumann, \emph{Mathematical Foundations of Quantum
Mechanics}, Princeton University Press (1955).

\bibitem {partovi}M. H. Partovi, \emph{Quantum thermodynamics}, Phys. Lett.
\textbf{A137}, 440 (1989).

\bibitem {shannon}C. Shannon, \emph{A mathematical theory of communication},
Bell Syst. Tech. J. \textbf{27} (1948): 379-423, 623-655.

\bibitem {landauer2}R. Landauer, \emph{Comment on physical limits to quantum
flux parametron operation}, Physica \textbf{C208}, 205 (1993).

\bibitem {nielsen1}M. A. Nielsen, C. M. Caves, B. Schumacher, and H. Barnum,
\emph{Information-theoretic approach to quantum error correction and
reversible measurement}, \textit{Proc. R. Soc. Lond.} \textbf{A454}, 277 (1998).

\bibitem {nielsen2}M. A. Nielsen and I. L. Chuang, \emph{Quantum Computing and
Quantum Information}, Cambridge University Press (2000).

\bibitem {jc}E. T. Jaynes and F. W. Cummings, \emph{Comparison of quantum and
semiclassical radiation theories with application to the beam maser}, Proc.
IEEE \textbf{51}, 89 (1963).

\bibitem {ben}M. Ben-Or, D. Gottesman, and A. Hassidim, \emph{Quantum
refrigerator}, arXiv:quant-ph/1301.1995 (2013).

\bibitem {hwang}W. Y. Hwang, D. Ahn, and S. W. Hwang, \emph{Correlated errors
in quantum-error corrections}, Phys. Rev. \textbf{A63}, 022303 (2001).

\bibitem {leung}D. W. Leung, M. A. Nielsen, I. L. Chuang, and Y. Yamamoto,
\emph{Approximate quantum error correction can lead to better codes}, Phys.
Rev. \textbf{A56}, 2567 (1997).

\bibitem {nori}K. Maruyama, F. Nori, and V. Vedral, \emph{Colloquium: The
physics of Maxwell's demon and information}, Rev. Mod. Phys. \textbf{81}, 1 (2009).

\bibitem {hillery}J.\ A. Bergou, U. Herzog, and M. Hillery,
\emph{Discrimination of quantum states}, Lect. Notes Phys. \textbf{649}, 417 (2004).

\bibitem {tony}A. Chefles, \emph{Quantum state discrimination}, Contemporary
Physics \textbf{41}, 401 (2000).

\bibitem {helstrom}C. W. Helstrom, \emph{Quantum Detection and Estimation
Theory}, Academic Press, New York (1976).

\bibitem {dieks}D. Dieks, \emph{Overlap and distinguishability of quantum
states}, Phys. Lett. \textbf{A126}, 303 (1988).

\bibitem {hardy}J. Walgate, A. J. Short, L. Hardy, and V. Vedral, \emph{Local
distinguishability of multipartite orthogonal quantum states}, Phys. Rev.
Lett. \textbf{85}, 4972 (2000).

\bibitem {ueda}M. Koashi and M. Ueda, \emph{Reversing measurement and
probabilistic quantum error correction}, Phys. Rev. Lett. \textbf{82}, 2598 (1999).

\bibitem {jay-science}E. T. Jaynes,\emph{ Comments on an article by Ulric
Neisser}, Science \textbf{140}, 216 (1963).

\bibitem {ric}R. Feynman, \emph{Simulating physics with computers}, Int. J.
Theor. Phys. \textbf{21}, 467 (1982).

\bibitem {sb}S. Bandyopadhyay, \emph{Entanglement, mixedness, and perfect
local discrimination of orthogonal quantum states}, Phys. Rev. \textbf{A85},
042319 (2012).

\bibitem {jab}J. A. Bergou, E. Feldman, and M. Hillery,\emph{ Optimal
unambiguous discrimination of two subspaces as a case in mixed-state
discrimination}, Phys. Rev. \textbf{A73}, 032107 (2006).

\bibitem {damian}D. Markham, J. A. Miszczak, Z. Puchala, and K. Zyczkowski,
\emph{Quantum state discrimination: A geometric approach}, Phys. Rev.
\textbf{A77}, 042111 (2008).

\bibitem {prl}B. Terhal, D. P. DiVincenzo, and D. W. Leung, \emph{Hiding bits
in Bell states}, Phys. Rev. Lett. \textbf{86}, 5807 (2001).

\bibitem {jarom}J. Fiurasek and M. Jezek, \emph{Optimal discrimination of
mixed quantum states involving inconclusive results}, Phys. Rev. \textbf{A67},
012321 (2003).

\bibitem {kl}E. Knill and R. Laflamme, \emph{Theory of quantum
error-correcting codes}, Phys. Rev. \textbf{A55}, 900 (1997).

\bibitem {beny}C. Beny and O. Oreshkov, \emph{General conditions for
approximate quantum error correction and near-optimal recovery channels},
Phys. Rev. Lett. \textbf{104}, 120501 (2010).
\end{thebibliography}
\end{document}